\documentclass[a4paper,11pt]{article}
\pdfoutput=1

\usepackage{jcappub} % for details on the use of the package, please
\usepackage[T1]{fontenc} % if needed
\usepackage{graphicx}
\usepackage{epstopdf}

\title{Heating of galactic gas by dark matter annihilation in ultracompact minihalos}

\author[a]{Hamish A. Clark,}
\author[a]{Nikolas Iwanus,}
\author[b]{Pascal J. Elahi,}
\author[a]{Geraint F. Lewis}
\author[c]{and Pat Scott}

\affiliation[a]{Sydney Institute for Astronomy, School of Physics A28, The University of Sydney, NSW 2006, Australia}
\affiliation[b]{International Centre for Radio Astronomy Research, University of Western Australia, 35 Stirling Highway, Crawley, WA 6009, Australia}
\affiliation[c]{Department of Physics, Imperial College London, Blackett Laboratory, Prince Consort Road, London SW7 2AZ, UK}

\emailAdd{hamish.clark@sydney.edu.au}
\emailAdd{nikolas.iwanus@sydney.edu.au}
\emailAdd{pascal.elahi@uwa.edu.au}
\emailAdd{geraint.lewis@sydney.edu.au}
\emailAdd{p.scott@imperial.ac.uk}

\abstract{The existence of substructure in halos of annihilating dark matter would be expected to substantially boost the rate at which annihilation occurs. Ultracompact minihalos of dark matter (UCMHs) are one of the more extreme examples of this.  The boosted annihilation can inject significant amounts of energy into the gas of a galaxy over its lifetime. Here we determine the impact of the boost factor from UCMH substructure on the heating of galactic gas in a Milky Way-type galaxy, by means of N-body simulation.  If $1\%$ of the dark matter exists as UCMHs, the corresponding boost factor can be of order $10^5$. For reasonable values of the relevant parameters (annihilation cross section $3\times10^{-26} ~\textrm{cm}^3~ \textrm{s}^{-1}$, dark matter mass 100\,GeV, 10\% heating efficiency), we show that the presence of UCMHs at the 0.1\% level would inject enough energy to eject significant amounts of gas from the halo, potentially preventing star formation within $\sim$1\,kpc of the halo centre.}

\begin{document}
\maketitle
\flushbottom

\section{Introduction}
\label{sec:intro}
The nature of dark matter (DM) is one of the largest unresolved mysteries in modern astrophysics. Constituting approximately 80\% of the matter density of the Universe, its identity is as yet unknown. Weakly interacting massive particles (WIMPs) provide one of the most persuasive solutions, as the present-day abundance of dark matter is similar to that expected for a particle produced thermally in the early Universe via electroweak interactions \cite{Jungman1996,Bergstrom2000,Bertone2005,Bergstrom2009}. The annihilations of these WIMPs would produce energetic particles, such as neutrinos, electron-positron pairs, or gamma rays. The search for the particle nature of dark matter would be significantly aided if the energy injected into the Universe by these particles had an observable effect. 

The pervasive and persistent injection of energy due to the presence of annihilating dark matter is expected to affect star formation and evolution \cite{SalatiSilk89, BouquetSalati89a, Ascasibar2007, Moskalenko07, Bertone07, Spolyar08, Fairbairn08, Scott08a, Iocco08a, Iocco08b, Taoso08, Yoon08, Scott09, Casanellas09, Ripamonti2010, Zackrisson10a, Zackrisson10b, Scott11,Smith2012,Stacy2014}, galaxy formation \cite{Ripamonti2007,Natarajan2009,Wechakama2011,Schon2015}, and the ionisation history of the early Universe \cite{Ripamonti2007a,Natarajan2008,Cirelli2009,Natarajan2009a,Natarajan2010}. These effects may be difficult to detect if the canonical annihilation cross section is assumed ($\langle\sigma v\rangle = 3\times 10^{-26} \textrm{ cm}^3~\textrm{s}^{-1}$). However, processes that could potentially boost the rate of annihilation have been proposed, such as Sommerfeld enhancement \cite{Hisano05,AHDM,Shepherd09,Lattanzi2009,Hryczuk11}, resonant enhancement \cite{Ibe2009} and enhancement due to the existence of dark matter substructure \cite{Strigari2007,Kuhlen2008,Elahi2009,Scott2009,Kamionkowski2010,Anderhalden2013,Bartels2015,Stref2016}. Should these annihilation boosts occur simultaneously, they could combine to increase the effects of annihilation by several orders of magnitude.

Previously, calculation of the annihilation boost due to substructure has depended heavily upon the extrapolation of halo properties over a large range of mass and density scales (for a discussion of the uncertainty in these assumptions, see \cite{Mack2014}). For example, modelling substructure using the Navarro-Frenk-White (NFW) profile for subhalos requires substantial extrapolation of the mass-concentration relation and radial mass distribution seen in N-body simulations \cite{Sanchez-Conde2014}. Conversely, if a significant fraction of dark matter exists in ultracompact minihalos (UCMHs) \cite{Berezinsky2003,Ricotti2009} no such extrapolation is required, as these objects are not disrupted by gravitational interactions, and so their distribution would be expected to follow the `bulk' density of dark matter.

N-body simulations have been an indispensable tool in the study of large-scale structure and cosmology \cite{Bertschinger1985,1985davis,2005natureSpringel}. However, by their nature, such simulations always have a finite resolution --- below which there may be poorly-understood physics or unresolved substructure. In N-body studies involving dark matter annihilation, where the annihilation rate is proportional to the density squared, unresolved substructure represents a major problem. A `clumpy' distribution with the same average density as a `smooth' distribution would lead to a significantly larger annihilation rate. Although this issue could in principle be solved by increasing the resolution of the simulation, the computational costs become prohibitive, as the largest modern simulations have achieved resolutions only on the order of a few kpc.

UCMHs have been proposed as a form of high density dark matter structure \cite{Berezinsky2003,Ricotti2009,Scott2009,Bringmann2012,Berezinsky2012,Berezinsky2013}. Produced by large-amplitude overdensities ($\delta \gtrsim 10^{-3}$) in the early universe, these dense halos collapse shortly after matter-radiation equality. This early formation means that the dark matter collapses by almost pure radial infall, leading to a steep density profile ($\rho \propto r^{-9/4}$ \cite{Bertschinger1985}) compared to that of the `standard' NFW halo ($\rho \propto r^{-1}$ \cite{Navarro1996}). These extremely dense cores are expected to exhibit large amounts of dark matter annihilation, leading to substantial production of high-energy annihilation products \cite{Scott2009}. In the event that even a small fraction of dark matter is contained within UCMHs, a significant amount of energy can be released via annihilation, affecting both structure formation and the observable properties of the cosmic `dark ages' \cite{Zhang2011,Yang2016}.

In this paper we consider the annihilation boost factor provided by the presence of UCMH substructure. We provide an analytical form of the boost factor due to unresolved UCMHs as a function of the `smooth' local dark matter density. We then implement the boost in an N-body simulation of an idealized NFW halo, incorporating the energy injected by dark matter annihilation.  We use this simulation to determine the magnitude of the effect that such a boost would have on the heating of galactic gas. This is the first study to perform an N-body simulation including dark matter annihilation along with both dark matter and gas components, and is likewise the first step toward a full cosmological simulation.

\section{Dark Matter Ultracompact Minihalos}
Here we summarise the main background on UCMHs; more details can be found in Ref.\ \cite{Bringmann2012}. After matter-radiation equality, a UCMH has a radial density profile of
\begin{equation}
\rho_{\rm h}(r,z) = \kappa(z) r^{-\frac{9}{4}},
\label{UCMHrho}
\end{equation}
where
\begin{equation}
\kappa(z) = \frac{3f_\chi M_{\rm h}(z)}{16\pi R_{\rm h}(z)^\frac{3}{4}},
\end{equation}
and $f_\chi$ is the fraction of matter that is CDM, $M_h (z)$ is the mass of the halo at some redshift $z$, and $R_{\rm h}(z)$ is the effective radius of the halo. Here, the mass of the halo evolves from that at matter-radiation equality, $M_{\rm i}$, up to a redshift $z\sim10$, as
\begin{equation}
M_{\rm h}(z) = \left(\frac{z_{\rm eq}+1}{z+1}\right)M_{\rm i},
\end{equation}
after which star formation begins, and accretion on to the halo effectively ceases.

The effective radius of a UCMH, beyond which its DM density contrast $\delta < 2$, has numerically been found to be
\begin{equation}
\frac{R_{\rm h}(z)}{\rm pc} = 0.019 \left(\frac{1000}{z+1}\right)\left(\frac{M_{\rm h}(z)}{M_\odot}\right)^{\frac{1}{3}},
\label{UCMHradius}
\end{equation}
again plateauing at $z=10$.

Within the cusp of the halo, we truncate the dark matter density by considering the maximum possible remaining dark matter due to annihilation at some time $t$,
\begin{equation}
\rho_{\rm c, ann}(t) = \frac{m_\chi}{(t-t_i)\langle \sigma v\rangle},
\end{equation}
where $m_\chi$ is the particle mass of dark matter, $\langle \sigma v\rangle$ is its late-time thermally-averaged self-annihilation cross section, and $t_{\rm i}$ is the time at which annihilation first started taking place.  Conservatively, we take this as the time of matter-radiation equality (just before the halo collapsed), such that $t_{\rm i} = t(z_{\rm eq}) = 59~\textrm{Myr}$. From Eq.\ \ref{UCMHrho}, the radius of the annihilation core is then
\begin{equation}
r_{\rm c,ann} = \left(\frac{\kappa}{\rho_{\rm c}}\right)^{\frac{4}{9}}.
\end{equation}
However, in the case that annihilation does not dominate, the core size is determined by the angular momentum of the initial infalling gas, as follows:
\begin{equation}
\frac{r_{\rm c, ang}}{R_{\rm h}(z=0)} \approx 2.9 \times 10^{-7} \left(\frac{1000}{z_{\rm c} +1}\right)^{2.43}\left(\frac{M_{\rm h} (z=0)}{M_\odot}\right)^{-0.06},
\end{equation}
where $z_c$ is the redshift of latest collapse for a UCMH, taken to be $z_c = 1000$. We then take the radius of the core as the largest of these, i.e.
\begin{equation}
r_{\rm c} = \max{(r_{\rm c,ann}, r_{\rm c, ang})},
\label{UCMHcore}
\end{equation}
\begin{equation}
\rho_{\rm c} = \kappa r_{\rm c}^{-9/4},
\end{equation}
such that the full piecewise expression for the density of an ultracompact minihalo at some radius $r$ will be:
\begin{align}
\rho (0 \leq r \leq r_{\rm c}) &= \rho_{\rm c}, \label{piecewiseUCMH}\\
\rho (r_{\rm c} < r \leq R_{\rm h}) & = \kappa r^{-9/4}, \nonumber \\
\rho (r> R_{\rm h}) &= 0.\nonumber
\end{align}

\section{UCMH Annihilation Boost}
Assuming that the spatial distribution of UCMHs follows that of the bulk dark matter, we can define the fraction of DM contained within unresolved UCMH substructure as
\begin{equation}
f \equiv \frac{\Omega_{\rm UCMH}(M_{\rm fs} < M_{\rm h} < M_{\rm res})}{\Omega_{\rm CDM}},
\label{fdef}
\end{equation}
where $\Omega_{\rm CDM}$ is the density of cold dark matter, $\Omega_{\rm UCMH}$ is the density of UCMHs, $M_{\rm res}$ is the minimum numerically resolvable UCMH mass, and $M_{\rm fs}$ is the minimum halo mass allowed by free streaming of dark matter. Depending upon the exact model of dark matter taken, this can range from $10^{-9} ~M_\odot$ up to $10^{-1} ~M_\odot$ for UCMHs at $z=0$ \cite{Bringmann2009}.

Should numerically unresolvable UCMH substructure exist, the average annihilation rate will be increased by a `boost' factor, defined as
\begin{equation}
\mathcal{B}(f, \rho_\chi) \equiv \frac{A_{\rm sub} + A_{\rm smooth}}{A_0},
\end{equation}
where $A_{\rm sub}$ and $A_{\rm smooth}$ are the annihilation rate per unit volume due to substructure, and the remaining `smooth' DM component, respectively. $A_0$ is the rate in the case that no substructure is present (i.e. $f=0$).

The existence of substructure implies that the density of the smooth component is reduced by a factor of $1-f$. The annihilation rate per unit volume due to this remaining component may then be found as
\begin{equation}
A_{\rm smooth} = \frac{\langle \sigma v\rangle }{2m_\chi^2}(1-f)^2\rho_\chi^2,
\end{equation}
where $\rho_\chi$ is the {\it numerically resolved} local dark matter density.

For a spherically symmetric minihalo of radius $R_{\rm h}$ superimposed on a smooth background, the total DM annihilation rate due to substructure may be found as
\begin{equation}
\Phi(M_{\rm h}) = \frac{2\pi \langle \sigma v\rangle}{m_\chi^2}\int_{0}^{R_{\rm h}} \left(\rho_{\rm h}^2 + 2\left(1-f\right)\rho_{\rm h}\rho_\chi\right)r^2 dr.
\label{perhalo}
\end{equation}
Here we have included both substructure-substructure ($\rho_{\rm h}^2$) and substructure-background ($\rho_{\rm h}\rho_\chi$) annihilations, but neglected the $\rho_\chi^2$ term to avoid double counting self-annihilation of the smooth background component. The average annihilation rate per unit volume due to substructure is then
\begin{equation}
A_{\rm sub} = \int_{M_{\rm fs}}^{M_{\rm res}}\Phi(M_{\rm h})\frac{dn}{dM_{\rm h}} dM_{\rm h},
\label{Asub1}
\end{equation}
where $dn/dM_{\rm h}$ is the local UCMH mass function, which describes the differential number density of halos of mass $M_{\rm h}$, expressed per unit halo mass.  Expressing this instead in terms of the differential fraction of dark matter in UCMHs of mass $M_h$, we find
\begin{equation}
A_{\rm sub} = \frac{\rho_\chi}{f_\chi M_{\rm h}}\int_{M_{\rm fs}}^{M_{\rm res}}\Phi(M_{\rm h})\frac{df}{dM_{\rm h}} dM_{\rm h}.
\label{Asub2}
\end{equation}

In the case of UCMH substructure, substituting Eq.\ \ref{piecewiseUCMH} into Eq.\ \ref{perhalo} results in an expression of the form $\Phi \propto M_{\rm h}^k$, where $k\approx 1$. This allows us to define a useful density scale
\begin{equation}
\xi \equiv \frac{2m_\chi^2 \Phi}{\langle \sigma v \rangle f_\chi M_{\rm h}},
\label{xidef}
\end{equation}
which is approximately independent of halo mass. Evaluating Eq.\ \ref{perhalo} and substituting into Eq.\ \ref{xidef} provides a functional form for $\xi$ of
\begin{equation}
\xi(f,\rho_\chi) = \frac{4 \pi}{f_\chi M_{\rm h}}\left[\kappa^2\left(r_{\rm c}^{-3/2} - \frac{2}{3}R_{\rm h}^{-3/2}\right) + 2\kappa(1-f)\rho_\chi\left(\frac{4}{3}R_{\rm h}^{3/4} - r_{\rm c}^{3/4}\right)\right].
\end{equation}

The lack of dependence of this function on $M_{\rm h}$ allows us to simplify Eq.\ \ref{Asub2} even further, taking all but $df/dM_{\rm h}$ outside of the integral to give
\begin{equation}
\label{asub}
A_{\rm sub} = \frac{\langle \sigma v \rangle}{2m_\chi^2}f \rho_\chi\xi.
\end{equation}
This leads to a boost factor from unresolved UCMH substructure of
\begin{equation}
\mathcal{B}(f, \rho_\chi) = \frac{f\xi}{\rho_\chi} + (1-f)^2.
\label{eq:boost_factor}
\end{equation}

\begin{figure}
\centering
\includegraphics[width=0.49\textwidth]{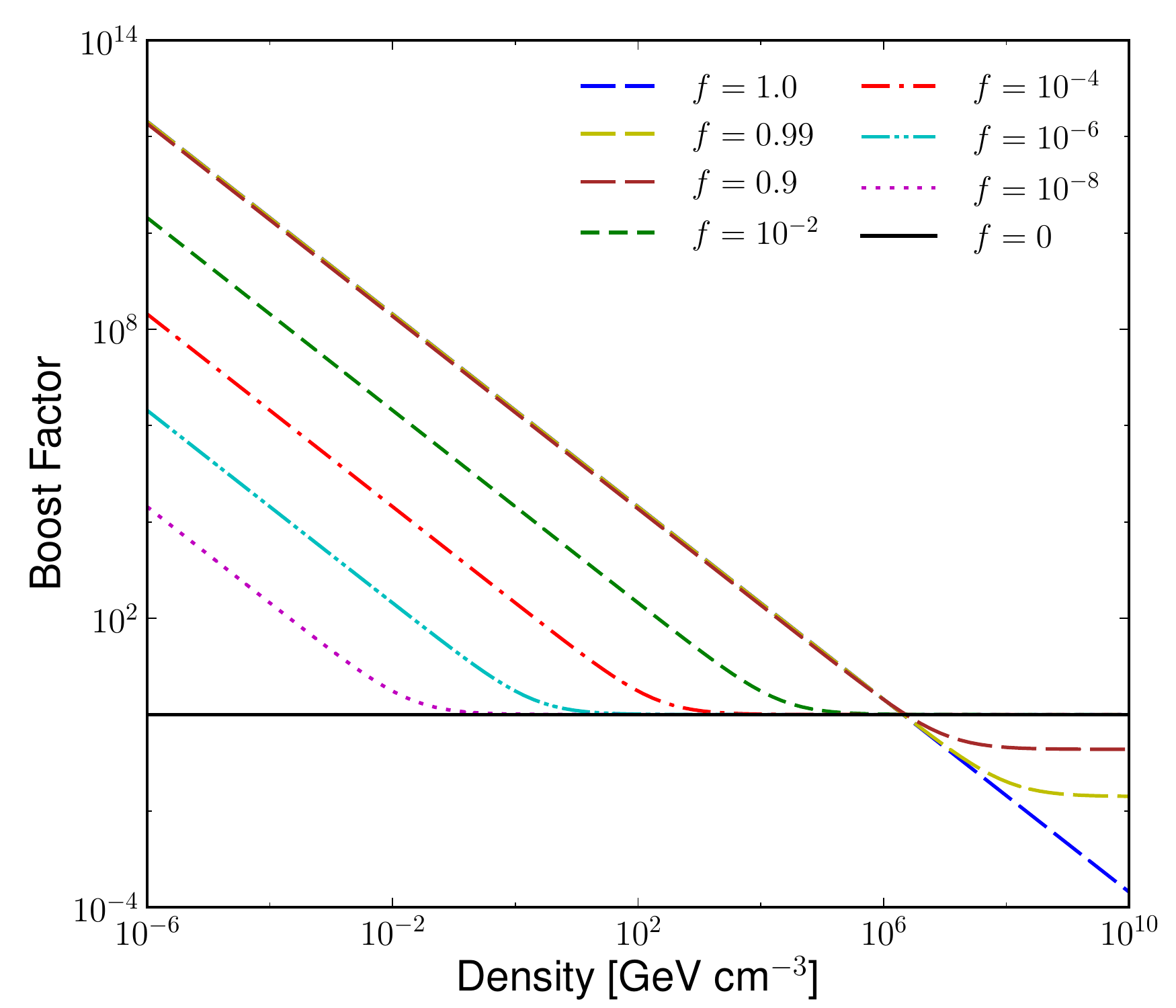}
\includegraphics[width=0.49\textwidth]{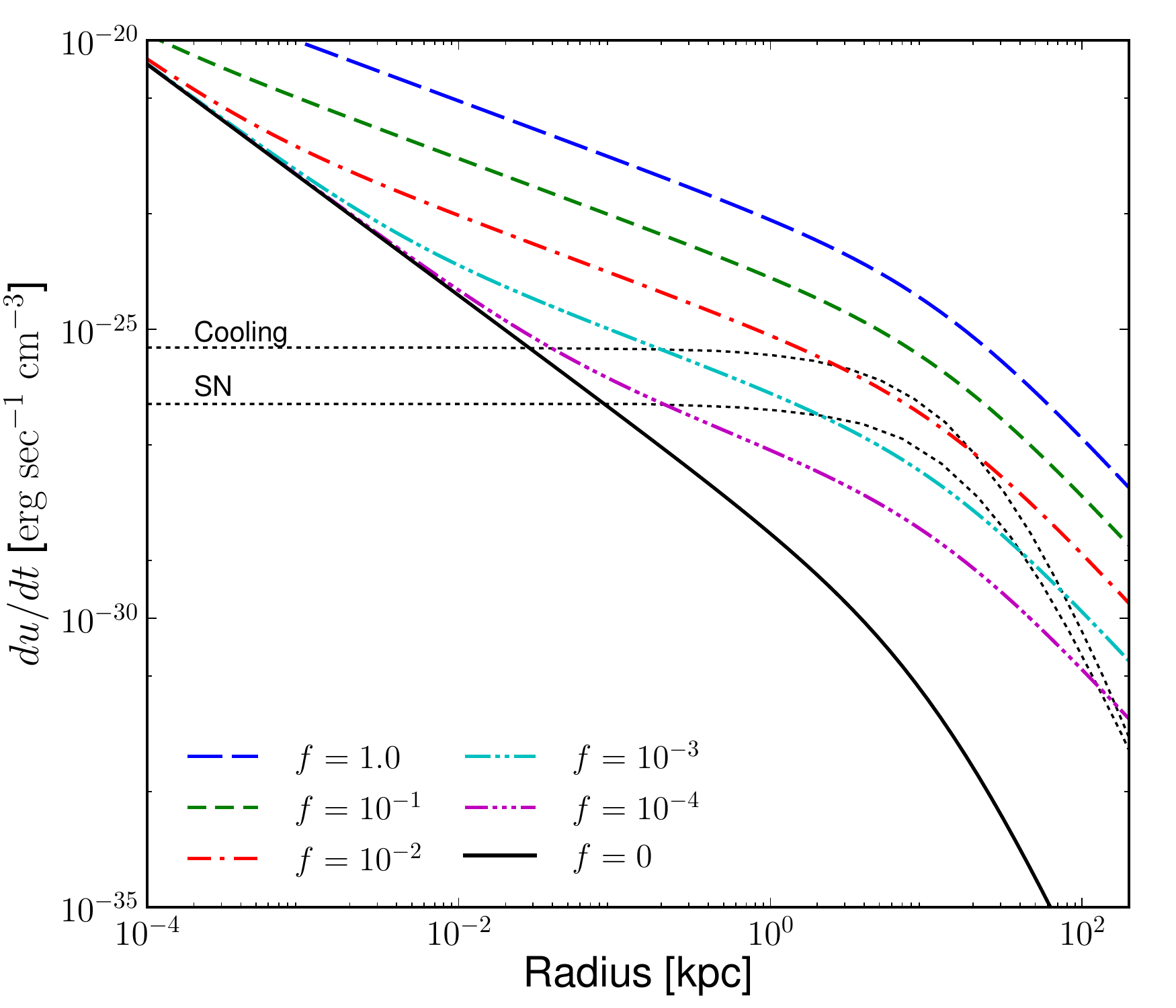}
\caption{\textit{Left}: The boost factor due to UCMH substructure as a function of the local resolved dark matter density, $\rho_\chi$. \textit{Right}: Rate of energy emitted by annihilation per unit volume as a function of radial distance from centre of a Milky Way-like dark matter halo, assuming $m_\chi = 100 \textrm{ GeV}$, $\langle\sigma v\rangle = 3\times 10^{-26} \textrm{ cm}^3 \textrm{ s}^{-1}$.  This halo has mass $10^{12}~M_\odot$, concentration parameter $c = 15$, and its density fixed to the local dark matter density $\rho_\chi = 0.4 \textrm{ GeV cm}^{-3}$ at the Sun's Galactic radius $r_\odot = 8.5 \textrm{ kpc}$. For comparison, we have plotted the supernova feedback and cooling rates (dotted lines) from \cite{Ascasibar2007}. Note that the typical radius of a Milky Way-like gas disk is of order 20--30 kpc, outside of which the effect of this heating due to annihilation should not be observable.}
\label{analytical_boost}
\end{figure}

This UCMH boost factor is entirely independent of the mass function $dn/dM_{\rm h}$, which is poorly constrained by observations for the low masses we consider here. It has also been suggested that UCMHs would track the bulk distribution of dark matter, as they are not significantly affected by tidal disruption \cite{Berezinsky2006, Berezinsky2008}. These two factors allow us to determine the overall annihilation rate per unit volume for a given $f$ purely from $\rho_\chi$, the total dark matter density resolved in N-body simulations.  We show the resulting relationship between $f$, $\rho_\chi$ and $\mathcal{B}$ in the left panel of Fig.\ \ref{analytical_boost}. As the local dark matter density increases, the boost factor decreases, ultimately providing an annihilation suppression in regions of very high ambient density. This is because the annihilation rate within the smooth component is proportional to $\rho_\chi^2$, whereas the flux from the UCMH component is proportional to $\Phi \rho_\chi$ (Eq.\ \ref{asub}). Therefore, if a significant fraction of dark matter is contained within substructure it will annihilate less efficiently when $\rho_\chi > \xi$ than if there were no substructure.

To understand the impacts of such boosts under actual Galactic conditions, we calculate the total boosted emission as $\mathcal{B}A_0$ --- assuming a density profile of a Milky Way-like NFW halo \cite{Navarro1996}. In the right panel of Fig.\ \ref{analytical_boost} we show the rate of energy emission from annihilation per unit volume, as a function of radius from the centre of a Milky Way-like halo. The density-dependent boost provides a much flatter Galactic annihilation profile than smooth DM alone, injecting a significantly larger amount of energy in the outer regions of the halo. Comparing to the local cooling rate of the gas, we see that for $f\gtrsim 10^{-2}$, the energy emitted by dark matter annihilation exceeds the cooling rate.\footnote{Cooling of the gas is here due to processes that produce photons: bremsstrahlung, collisional ionization, recombination, and collisional excitation \cite{Sutherland1993}. These photons free stream out of the galaxy, removing energy from the gas.} The rate of star formation would be expected to be affected, depending on how much of the annihilation energy actually goes into heating the interstellar medium. For the case of $f\lesssim 10^{-2}$, the heating and cooling are balanced at radii that can be a significant fraction of the virial radius of the halo. 

Previously, non-observation of UCMH gamma-ray emission with \textit{Fermi}-LAT has been used to provide upper limits on $f$ \cite{Bringmann2012}. While these constraints are the strongest to date, they only apply directly to scenarios where UCMHs are all of the same mass (although limits on mass spectra can be obtained by integrating over the appropriate mass window). Given that the boost factor derived here applies equally to any mass distribution, we compare to the weakest such limit of any mass: $f\lesssim 10^{-3}$, provided by diffuse emission within the Galaxy. If even 0.1\% of the dark matter in the Milky Way is contained within UCMHs, the energy emitted can be substantial. If e.g.\ 10\% of this energy were absorbed by the local gas, the heating caused would be sufficient to quench star formation within the inner few tens of pc of the Galactic centre, and would be comparable to that from supernovae within a radius of a few hundred pc.

\section{Idealised N-body Simulation}
To investigate how significant the heating by boosted annihilation from UCMH substructure could be in a real galaxy, we added the energy injection to an N-body simulation.  For this we used new modules written for the cosmological N-body code \textsf{Gadget-2} \cite{Springel2005a}, designed to calculate the heat injected into gas particles by absorption of the local annihilation products from arbitrary dark matter distributions (Iwanus et al. in prep).  These modules use smooth particle hydrodynamics (SPH) to estimate the dark matter density at the location of every gas particle, which they then use to determine the annihilation energy to be injected into each gas particle.

Considerable energy injection into baryonic matter would provide an increased pressure within galaxies, forcing gas out of their centre and altering their structure.  We consider the case of dark matter annihilation boosted by the factor given in Eq.\ \ref{eq:boost_factor}, resulting in an energy absorption rate per unit volume of
\begin{equation}
   \frac{du}{dt} = \varepsilon\mathcal{B}(f,\rho_{\chi}) \frac{\langle\sigma v \rangle}{m_{\chi}}\rho_{\chi}^{2}.
	\label{eq:gas_injection}
\end{equation}
Here $\varepsilon$ is the fraction of the energy released per annihilation absorbed by the local gas. In the galactic regime, the electron-positron  annihilation channel would provide one of the largest absorption fractions. These charged particles in a dense galactic environment would be expected to both produce synchrotron radiation and undergo inverse Compton scattering as the dominating sources of energy loss -- effects that have been used recently to search for evidence of dark matter annihilation at the Galactic Centre \cite{Hooper2007,Hooper2011,Cholis2015}. The authors of Ref. \cite{Delahaye2010} accounted for these dominating processes in a study of the absorption of cosmic rays within the Milky Way.  At typical galactic densities, they found that $\gtrsim$50\% of the energy of a 100 GeV positron is deposited over a length scale of 1\,kpc. Given the uncertainties in this value for generalized galactic regimes, we explore the assumption of $\varepsilon = 0.01$, as well as the less conservative $\varepsilon = 0.1$. Future work is needed to better approximate this value.\footnote{Previous investigation of this efficiency has been undertaken either in homogeneous cosmological regimes \cite{SPF, Slatyer2013}, or for low-mass high-redshift halos \cite{Schon2015}, providing an estimate ranging from $\varepsilon = 1$ down to $\varepsilon = 0.01$.}

By applying our boost factor as a form of subgrid physics we have implicitly assumed that all of the UCMHs are numerically unresolved. By increasing the resolution of a simulation, the gravitational effects of large-mass UCMHs would become apparent. In this case care must be taken that the UCMH fraction $f$ only counts those with mass $< M_{ \rm res}$, while the larger, resolvable UCMHs would have to be explicitly placed into the initial conditions of the simulation. In what follows, we have not explicitly placed any UCMHs into the simulation, rather injecting the effect of UCMH annihilation with Eq. \ref{eq:gas_injection}.

Using \textsf{GalactICs} \cite{galactics1, galactics2, galactics3}, we generated an NFW halo of mass $10^{12}\,M_{\odot}$ and concentration $c=15$, consisting of $10^5$ particles. We converted $\Omega_{\rm b}/\Omega_{\rm CDM} \approx 20\%$ of the dark matter into gas particles of the same mass, with a thermal energy equal to their local velocity dispersion. We also reduced the kinetic energies of the gas particles throughout the entire simulation so as to maintain energy conservation, and evolved the halo for 15\,Gyr, to ensure that it had fully virialised. We initiated dark matter self-annihilation with a cross section of $\langle \sigma v\rangle = 3 \times 10^{-26} ~\textrm{cm}^3~\textrm{s}^{-1}$ and a particle mass $m_\chi = 100~\textrm{GeV}$, and evolved the simulation for a further 5\,Gyr using our modified version of \textsf{Gadget-2}.  We repeated the final step for a number of different UCMH fractions and heating efficiencies. While the assumption of an idealised NFW halo will underestimate the overall annihilation due to the lack of `natural' dark matter structure, it allows us to investigate the effect that the introduction of UCMH substructure will make.

\begin{figure}
\centering
\includegraphics[width=\textwidth]{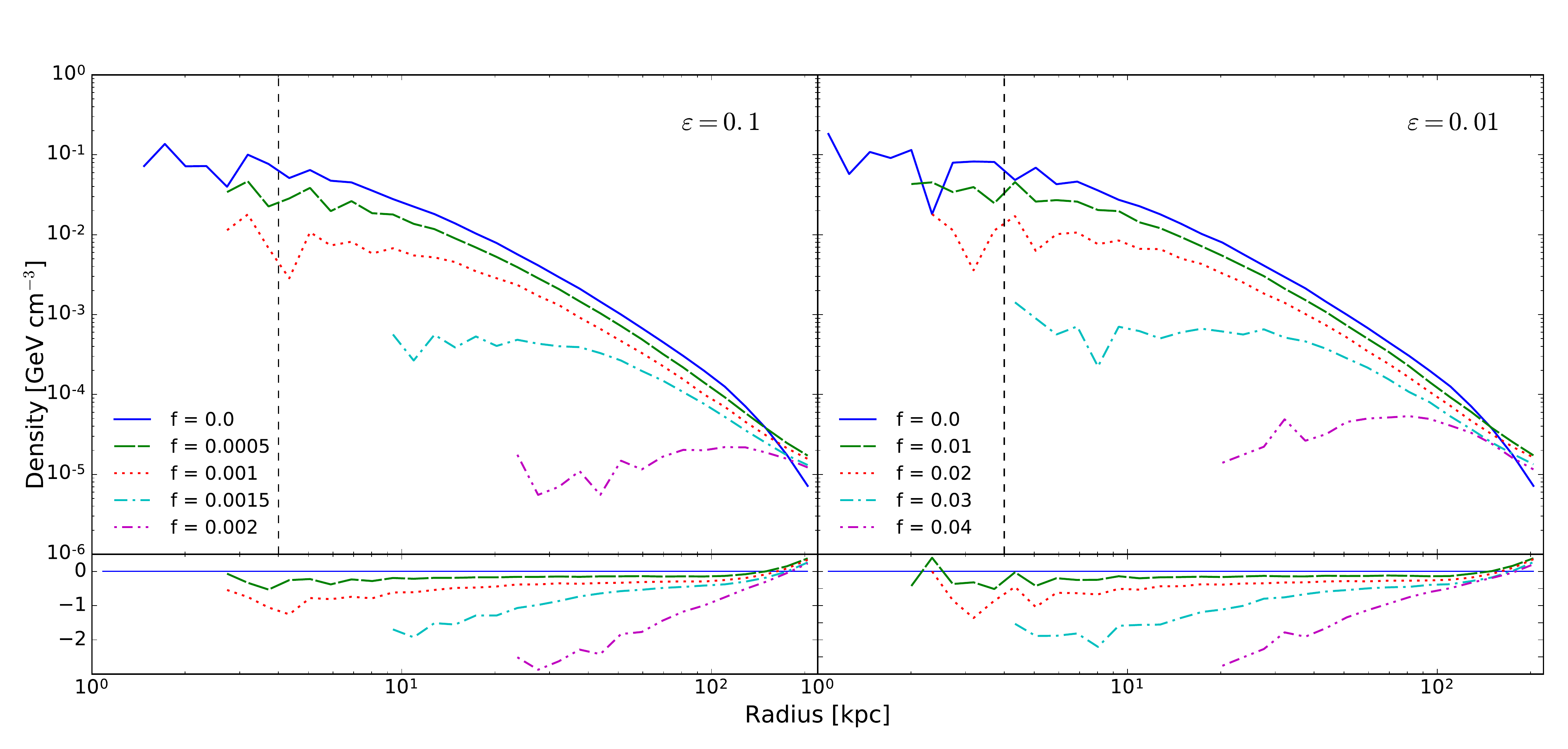}
\caption{ The radial profiles of gas in a Milky Way-like halo simulated with annihilation in UCMH substructures. Here we have assumed that 10\% (\textit{left}) and 1\% (\textit{right}) of the annihilation energy goes into heating of the gas, and a dark matter model with $\langle \sigma v\rangle = 3 \times 10^{-26} ~\textrm{cm}^3~\textrm{s}^{-1}$ and $m_\chi = 100~\textrm{GeV}$. For comparison to the case without substructure ($f=0$), we have plotted the logarithmic residuals below each figure. The vertical dashed line represents the effective resolution of our simulation ($r \approx 10$ kpc) --- for \textsc{Gadget}, this corresponds to 2.7 times the gravitational softening length. Image prepared with \textsf{pynbody} \cite{pynbody}.}
\label{fig:profile_sim}
\end{figure}

We present the resulting gas density profiles in Fig.\ \ref{fig:profile_sim}. It can be seen that as the substructure fraction increases, more gas is ejected from the halo due to slow, persistent heating by annihilation. We see that smooth dark matter ($f=0$) does not impart significant energy to the gas, but even a tiny fraction of UCMHs can produce an appreciable change in the gas profile at regions of high dark matter density. In the case of very large substructure fractions we find very significant gas outflow.  In these cases, the majority of the gas is removed from the galaxy.

Disturbance of the gas is observed for substructure fractions as low as $f=1\times 10^{-2}$, even under the most conservative assumptions. Constraints on the UCMH fraction at the time of matter-radiation equality have been determined from the effect of DM annihilation on the integrated optical depth of the CMB \cite{Zhang2011}. However, when extended to the present day these constraints weaken to the point of saturation ($f\leq 1$). By contrast, the strong constraints placed upon $f$ with gamma-ray searches by \textit{Fermi}-LAT \citep{Bringmann2012} have excluded such large UCMH abundances ($f\lesssim 10^{-3}$). This strong limit rules out any effect for an assumed heating efficiency of $\varepsilon = 0.01$. A slightly less conservative estimate of $\varepsilon = 0.1$ yields significant gas heating for $f = 5 \times 10^{-4}$, which is well within observational limits.

\begin{figure}
\centering
\includegraphics[width=0.49\textwidth]{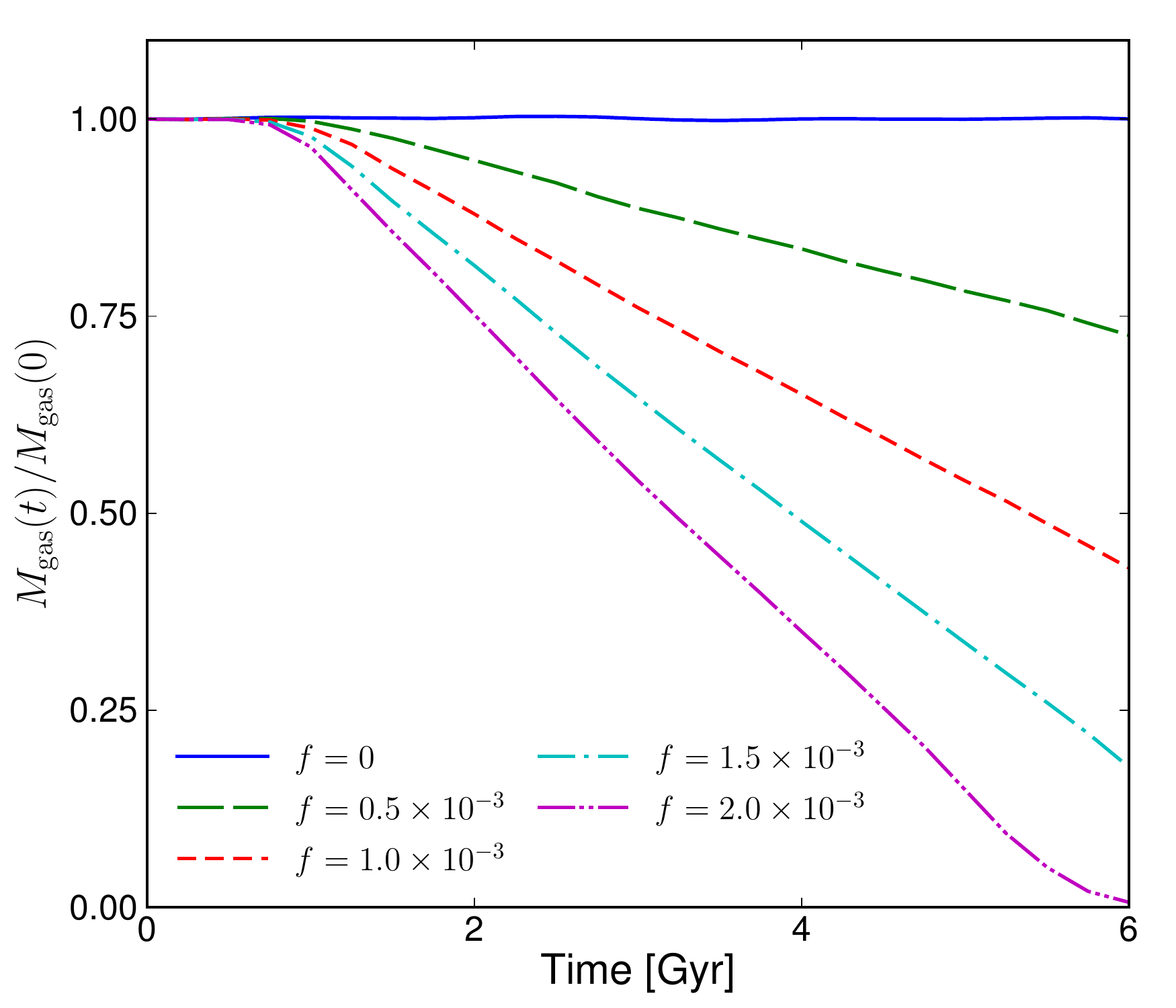}
\includegraphics[width=0.49\textwidth]{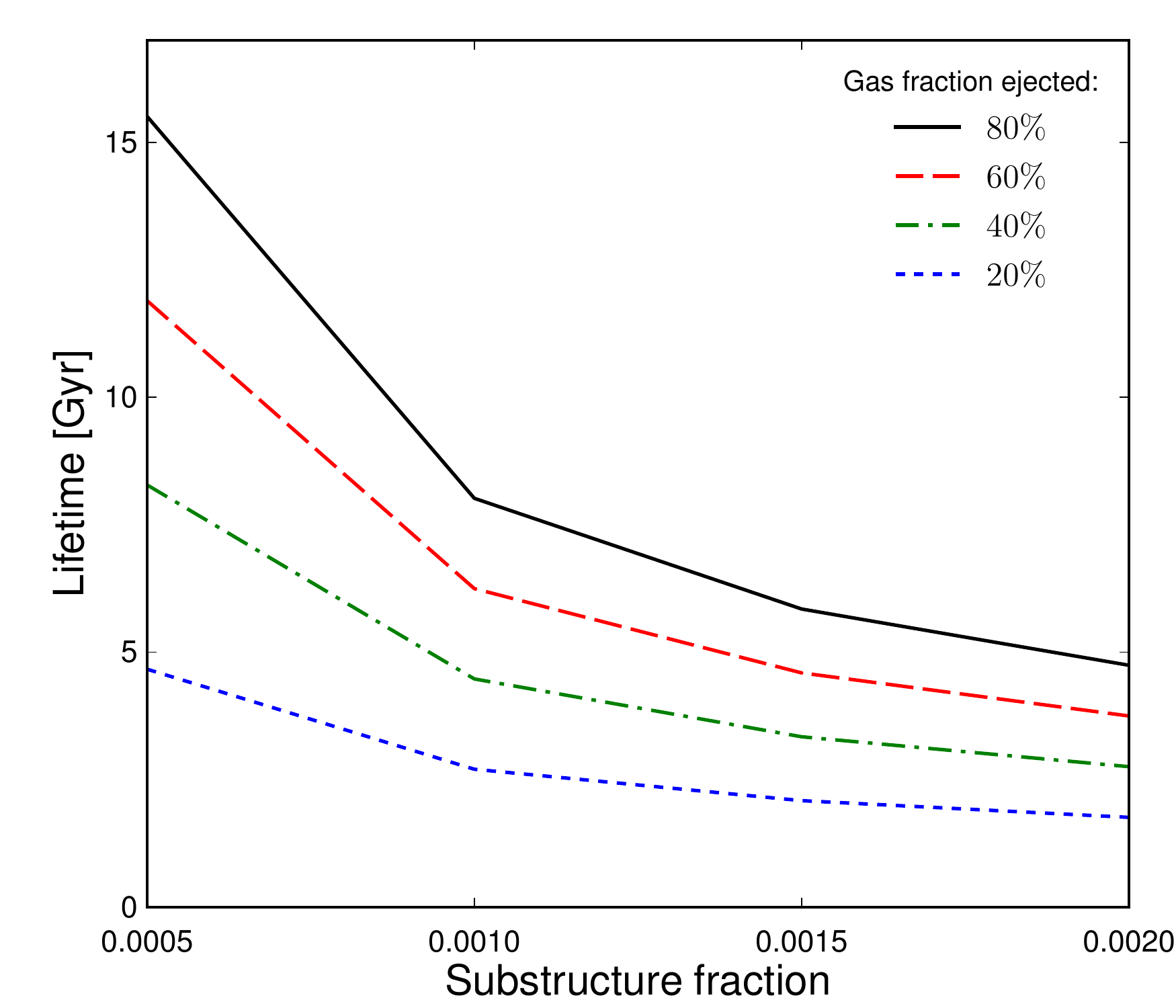}
\caption{\textit{Left}: The mass of the gas the halo enclosed within $R_{200}$ as a function of time, for a range of substructure fractions. Image prepared with \textsf{pynbody} \cite{pynbody}. \textit{Right}: The time taken to eject a given fraction of the halo's gas by annihilation, as a function of substructure abundance. Parameters as per Fig.\ \ref{fig:profile_sim}, with $\varepsilon = 0.1$.}
\label{fig:mass_vs_time}
\end{figure}

To further understand the evolution of the halo's gas, it is useful to trace its mass as a function of time. We therefore integrated the gas density inward from $R_{200}$, the radius at which the average density within is equal to 200 times the critical density of the Universe. We display this mass as a function of time for a range of substructure fractions in Fig.\ \ref{fig:mass_vs_time} (left). We can see that once the annihilation is initialised at $t=0$, high substructure fractions ($f\gtrsim 10^{-3}$) cause significant mass to be ejected from the halo, whereas very little gas is ejected if there is no substructure ($f=0$). We additionally characterise this heating by defining a `lifetime' of the halo. We define this as the time taken to reduce the gas content of a halo by a given percentage, and show it in Fig.\ \ref{fig:mass_vs_time} (right). We find that as the substructure fraction increases, this lifetime decreases rapidly, converging for $f \gtrsim 0.002$ to less than 5 Gyr.

\section{Conclusion}
In this paper we have investigated the effect of a population of unresolved UCMHs on the energy injected into a galaxy by dark matter self-annihilation. We showed that the increase in annihilation rate per unit volume due to UCMH substructure is independent of subhalo mass, and thus may be mapped as a function of the overall dark matter density alone. The elegance and utility of this finding may be seen when it is applied to N-body simulations, within which the physics beneath the resolution of the simulation is notoriously difficult to model. As the annihilation boost factor due to UCMH substructure depends only upon the large-scale properties of the density field, the boost factor may be easily incorporated into any N-body simulation or indeed any investigation of the annihilation of dark matter.

By application to such an N-body simulation of an idealized halo, we found that the boost factor due to the presence of UCMH substructure can produce a significant change in the distribution of its gas. If 10\% of the energy from annihilation is absorbed by the gas, and more than $\sim$$0.05\%$ of dark matter in the halo is contained within UCMHs, its centre would be evacuated of a significant fraction of its gas after only 5\,Gyr. In the future, should a full cosmological N-body simulation be undertaken including the presence of UCMHs, such a large amount of injected energy may need to be countered by other mechanisms in order to retain solutions that resemble observations. This may be done by the addition of increased cooling mechanisms, or by recalibration of feedback (from e.g.\ active galactic nuclei, star formation or black hole accretion). If these processes were to become better constrained from the observational side, this could be used as a method for determining an upper limit to the fraction of dark matter in the form of UCMHs. Conversely, observations of the CMB could be compared to such simulations to place a limit on the abundance of UCMHs at early times. Given the direct link between UCMHs and the properties of the early Universe \cite{JG10,Bringmann2012, Shandera12,Clark2016,Aslanyan16}, such a constraint would be of substantial cosmological importance.

\acknowledgments
The authors acknowledge the University of Sydney HPC service for providing computational resources that have contributed to the research results reported within this paper. HAC and NI acknowledge the Australian Postgraduate Awards (APA), through which this work was financially supported. PS is supported by STFC (ST/K00414X/1 and ST/N000838/1).


\bibliographystyle{JHEP}
\bibliography{boost}

\providecommand{\href}[2]{#2}\begingroup\raggedright\begin{thebibliography}{10}

\bibitem{Jungman1996}
G.~{Jungman}, M.~{Kamionkowski} and K.~{Griest}, \emph{{Supersymmetric dark
  matter}}, \href{http://dx.doi.org/10.1016/0370-1573(95)00058-5}{\emph{Phys.
  Rep.} {\bf 267} (Mar., 1996) 195--373},
  [\href{https://arxiv.org/abs/hep-ph/9506380}{{\tt hep-ph/9506380}}].

\bibitem{Bergstrom2000}
L.~{Bergstr{\"o}m}, \emph{{Non-baryonic dark matter: observational evidence and
  detection methods}},
  \href{http://dx.doi.org/10.1088/0034-4885/63/5/2r3}{\emph{Rep. Prog. Phys.}
  {\bf 63} (May, 2000) 793--841},
  [\href{https://arxiv.org/abs/hep-ph/0002126}{{\tt hep-ph/0002126}}].

\bibitem{Bertone2005}
G.~{Bertone}, D.~{Hooper} and J.~{Silk}, \emph{{Particle dark matter: evidence,
  candidates and constraints}},
  \href{http://dx.doi.org/10.1016/j.physrep.2004.08.031}{\emph{Phys. Rep.} {\bf
  405} (Jan., 2005) 279--390},
  [\href{https://arxiv.org/abs/hep-ph/0404175}{{\tt hep-ph/0404175}}].

\bibitem{Bergstrom2009}
L.~{Bergstr{\"o}m}, \emph{{Dark matter candidates}},
  \href{http://dx.doi.org/10.1088/1367-2630/11/10/105006}{\emph{New J. Phys.}
  {\bf 11} (Oct., 2009) 105006}, [\href{https://arxiv.org/abs/0903.4849}{{\tt
  0903.4849}}].

\bibitem{SalatiSilk89}
P.~{Salati} and J.~{Silk}, \emph{{A stellar probe of dark matter annihilation
  in galactic nuclei}}, \href{http://dx.doi.org/10.1086/167177}{\emph{\apj}
  {\bf 338} (1989) 24--31}.

\bibitem{BouquetSalati89a}
A.~{Bouquet} and P.~{Salati}, \emph{{Life and death of cosmions in stars}},
  {\emph{\aap} {\bf 217} (1989) 270--282}.

\bibitem{Ascasibar2007}
Y.~{Ascasibar}, \emph{{Effect of dark matter annihilation on gas cooling and
  star formation}},
  \href{http://dx.doi.org/10.1051/0004-6361:20066880}{\emph{A\&A} {\bf 462}
  (Feb., 2007) L65--L68}, [\href{https://arxiv.org/abs/astro-ph/0612130}{{\tt
  astro-ph/0612130}}].

\bibitem{Moskalenko07}
I.~V. {Moskalenko} and L.~L. {Wai}, \emph{{Dark Matter Burners}},
  \href{http://dx.doi.org/10.1086/516708}{\emph{\apjl} {\bf 659} (2007)
  L29--L32}, [\href{https://arxiv.org/abs/astro-ph/0702654}{{\tt
  astro-ph/0702654}}].

\bibitem{Bertone07}
G.~{Bertone} and M.~{Fairbairn}, \emph{{Compact stars as dark matter probes}},
  \href{http://dx.doi.org/10.1103/PhysRevD.77.043515}{\emph{\prd} {\bf 77}
  (Feb., 2008) 043515}, [\href{https://arxiv.org/abs/0709.1485}{{\tt
  0709.1485}}].

\bibitem{Spolyar08}
D.~{Spolyar}, K.~{Freese} and P.~{Gondolo}, \emph{{Dark Matter and the First
  Stars: A New Phase of Stellar Evolution}},
  \href{http://dx.doi.org/10.1103/PhysRevLett.100.051101}{\emph{\prl} {\bf 100}
  (Feb., 2008) 051101}, [\href{https://arxiv.org/abs/0705.0521}{{\tt
  0705.0521}}].

\bibitem{Fairbairn08}
M.~{Fairbairn}, P.~{Scott} and J.~{Edsj{\"o}}, \emph{{The zero age main
  sequence of WIMP burners}},
  \href{http://dx.doi.org/10.1103/PhysRevD.77.047301}{\emph{\prd} {\bf 77}
  (Feb., 2008) 047301}, [\href{https://arxiv.org/abs/{arXiv:0710.3396}}{{\tt
  {arXiv:0710.3396}}}].

\bibitem{Scott08a}
P.~{Scott}, J.~{Edsj{\"o}} and M.~{Fairbairn}, \emph{{Low mass stellar
  evolution with WIMP capture and annihilation}},  in \emph{Dark Matter in
  Astroparticle and Particle Physics: Dark 2007} (H.~K. {Klapdor-Kleingrothaus}
  and G.~F. {Lewis}, eds.), pp.~387--392, World Scientific, Singapore, 2008.
\newblock \href{https://arxiv.org/abs/{arXiv:0711.0991}}{{\tt
  {arXiv:0711.0991}}}.

\bibitem{Iocco08a}
F.~{Iocco}, \emph{{Dark Matter Capture and Annihilation on the First Stars:
  Preliminary Estimates}}, \href{http://dx.doi.org/10.1086/587959}{\emph{\apjl}
  {\bf 677} (Apr., 2008) L1--L4}, [\href{https://arxiv.org/abs/0802.0941}{{\tt
  0802.0941}}].

\bibitem{Iocco08b}
F.~{Iocco}, A.~{Bressan}, E.~{Ripamonti}, R.~{Schneider}, A.~{Ferrara} and
  P.~{Marigo}, \emph{{Dark matter annihilation effects on the first stars}},
  \href{http://dx.doi.org/10.1111/j.1365-2966.2008.13853.x}{\emph{\mnras} {\bf
  390} (Nov., 2008) 1655--1669}, [\href{https://arxiv.org/abs/0805.4016}{{\tt
  0805.4016}}].

\bibitem{Taoso08}
M.~{Taoso}, G.~{Bertone}, G.~{Meynet} and S.~{Ekstr{\"o}m}, \emph{{Dark matter
  annihilations in Population III stars}},
  \href{http://dx.doi.org/10.1103/PhysRevD.78.123510}{\emph{\prd} {\bf 78}
  (Dec., 2008) 123510}, [\href{https://arxiv.org/abs/0806.2681}{{\tt
  0806.2681}}].

\bibitem{Yoon08}
S.-C. {Yoon}, F.~{Iocco} and S.~{Akiyama}, \emph{{Evolution of the First Stars
  with Dark Matter Burning}},
  \href{http://dx.doi.org/10.1086/593976}{\emph{\apjl} {\bf 688} (Nov., 2008)
  L1--L4}, [\href{https://arxiv.org/abs/0806.2662}{{\tt 0806.2662}}].

\bibitem{Scott09}
P.~{Scott}, M.~{Fairbairn} and J.~{Edsj{\"o}}, \emph{{Dark stars at the
  Galactic Centre - the main sequence}},
  \href{http://dx.doi.org/10.1111/j.1365-2966.2008.14282.x}{\emph{\mnras} {\bf
  394} (2009) 82--104}, [\href{https://arxiv.org/abs/{arXiv:0809.1871}}{{\tt
  {arXiv:0809.1871}}}].

\bibitem{Casanellas09}
J.~{Casanellas} and I.~{Lopes}, \emph{{The Formation and Evolution of Young
  Low-mass Stars within Halos with High Concentration of Dark Matter
  Particles}}, \href{http://dx.doi.org/10.1088/0004-637X/705/1/135}{\emph{\apj}
  {\bf 705} (Nov., 2009) 135--143},
  [\href{https://arxiv.org/abs/0909.1971}{{\tt 0909.1971}}].

\bibitem{Ripamonti2010}
E.~{Ripamonti}, F.~{Iocco}, A.~{Ferrara}, R.~{Schneider}, A.~{Bressan} and
  P.~{Marigo}, \emph{{First star formation with dark matter annihilation}},
  \href{http://dx.doi.org/10.1111/j.1365-2966.2010.16854.x}{\emph{MNRAS} {\bf
  406} (Aug., 2010) 2605--2615}, [\href{https://arxiv.org/abs/1003.0676}{{\tt
  1003.0676}}].

\bibitem{Zackrisson10a}
E.~{Zackrisson}, P.~{Scott}, C.-E. {Rydberg}, F.~{Iocco}, B.~{Edvardsson},
  G.~{{\"O}stlin} et~al., \emph{{Finding High-redshift Dark Stars with the
  James Webb Space Telescope}},
  \href{http://dx.doi.org/10.1088/0004-637X/717/1/257}{\emph{\apj} {\bf 717}
  (July, 2010) 257--267}, [\href{https://arxiv.org/abs/{arXiv:1002.3368}}{{\tt
  {arXiv:1002.3368}}}].

\bibitem{Zackrisson10b}
E.~{Zackrisson}, P.~{Scott}, C.-E. {Rydberg}, F.~{Iocco}, S.~{Sivertsson},
  G.~{{\"O}stlin} et~al., \emph{{Observational constraints on supermassive dark
  stars}},
  \href{http://dx.doi.org/10.1111/j.1745-3933.2010.00908.x}{\emph{\mnras} {\bf
  407} (Sept., 2010) L74--L78},
  [\href{https://arxiv.org/abs/{arXiv:1006.0481}}{{\tt {arXiv:1006.0481}}}].

\bibitem{Scott11}
P.~{Scott}, A.~{Venkatesan}, E.~{Roebber}, P.~{Gondolo}, E.~{Pierpaoli} and
  G.~{Holder}, \emph{{Impacts of Dark Stars on Reionization and Signatures in
  the Cosmic Microwave Background}}, {\emph{\apj} {\bf 742} (July, 2011) 129},
  [\href{https://arxiv.org/abs/1107.1714}{{\tt 1107.1714}}].

\bibitem{Smith2012}
R.~J. {Smith}, F.~{Iocco}, S.~C.~O. {Glover}, D.~R.~G. {Schleicher}, R.~S.
  {Klessen}, S.~{Hirano} et~al., \emph{{Weakly Interacting Massive Particle
  Dark Matter and First Stars: Suppression of Fragmentation in Primordial Star
  Formation}}, \href{http://dx.doi.org/10.1088/0004-637X/761/2/154}{\emph{ApJ}
  {\bf 761} (Dec., 2012) 154}, [\href{https://arxiv.org/abs/1210.1582}{{\tt
  1210.1582}}].

\bibitem{Stacy2014}
A.~{Stacy}, A.~H. {Pawlik}, V.~{Bromm} and A.~{Loeb}, \emph{{The mutual
  interaction between Population III stars and self-annihilating dark matter}},
  \href{http://dx.doi.org/10.1093/mnras/stu621}{\emph{\mnras} {\bf 441} (June,
  2014) 822--836}, [\href{https://arxiv.org/abs/1312.3117}{{\tt 1312.3117}}].

\bibitem{Ripamonti2007}
E.~{Ripamonti}, M.~{Mapelli} and A.~{Ferrara}, \emph{{The impact of dark matter
  decays and annihilations on the formation of the first structures}},
  \href{http://dx.doi.org/10.1111/j.1365-2966.2006.11402.x}{\emph{MNRAS} {\bf
  375} (Mar., 2007) 1399--1408},
  [\href{https://arxiv.org/abs/astro-ph/0606483}{{\tt astro-ph/0606483}}].

\bibitem{Natarajan2009}
A.~{Natarajan}, J.~C. {Tan} and B.~W. {O'Shea}, \emph{{Dark Matter Annihilation
  and Primordial Star Formation}},
  \href{http://dx.doi.org/10.1088/0004-637X/692/1/574}{\emph{ApJ} {\bf 692}
  (Feb., 2009) 574--583}, [\href{https://arxiv.org/abs/0807.3769}{{\tt
  0807.3769}}].

\bibitem{Wechakama2011}
M.~{Wechakama} and Y.~{Ascasibar}, \emph{{Pressure from dark matter
  annihilation and the rotation curve of spiral galaxies}},
  \href{http://dx.doi.org/10.1111/j.1365-2966.2011.18275.x}{\emph{MNRAS} {\bf
  413} (May, 2011) 1991--2003}, [\href{https://arxiv.org/abs/1007.3179}{{\tt
  1007.3179}}].

\bibitem{Schon2015}
S.~{Sch{\"o}n}, K.~J. {Mack}, C.~A. {Avram}, J.~S.~B. {Wyithe} and
  E.~{Barberio}, \emph{{Dark matter annihilation in the first galaxy haloes}},
  \href{http://dx.doi.org/10.1093/mnras/stv1056}{\emph{MNRAS} {\bf 451} (Aug.,
  2015) 2840--2850}, [\href{https://arxiv.org/abs/1411.3783}{{\tt 1411.3783}}].

\bibitem{Ripamonti2007a}
E.~{Ripamonti}, M.~{Mapelli} and A.~{Ferrara}, \emph{{Intergalactic medium
  heating by dark matter}},
  \href{http://dx.doi.org/10.1111/j.1365-2966.2006.11222.x}{\emph{\mnras} {\bf
  374} (Jan., 2007) 1067--1077},
  [\href{https://arxiv.org/abs/astro-ph/0606482}{{\tt astro-ph/0606482}}].

\bibitem{Natarajan2008}
A.~{Natarajan} and D.~J. {Schwarz}, \emph{{Effect of early dark matter halos on
  reionization}},
  \href{http://dx.doi.org/10.1103/PhysRevD.78.103524}{\emph{Phys. Rev. D} {\bf
  78} (Nov., 2008) 103524}, [\href{https://arxiv.org/abs/0805.3945}{{\tt
  0805.3945}}].

\bibitem{Cirelli2009}
M.~{Cirelli}, F.~{Iocco} and P.~{Panci}, \emph{{Constraints on Dark Matter
  annihilations from reionization and heating of the intergalactic gas}},
  \href{http://dx.doi.org/10.1088/1475-7516/2009/10/009}{\emph{J. Cosmol.
  Astropart. Phys.} {\bf 10} (Oct., 2009) 009},
  [\href{https://arxiv.org/abs/0907.0719}{{\tt 0907.0719}}].

\bibitem{Natarajan2009a}
A.~{Natarajan} and D.~J. {Schwarz}, \emph{{Dark matter annihilation and its
  effect on CMB and hydrogen 21cm observations}},
  \href{http://dx.doi.org/10.1103/PhysRevD.80.043529}{\emph{\prd} {\bf 80}
  (Aug., 2009) 043529}, [\href{https://arxiv.org/abs/0903.4485}{{\tt
  0903.4485}}].

\bibitem{Natarajan2010}
A.~{Natarajan} and D.~J. {Schwarz}, \emph{{Distinguishing standard reionization
  from dark matter models}},
  \href{http://dx.doi.org/10.1103/PhysRevD.81.123510}{\emph{\prd} {\bf 81}
  (June, 2010) 123510}, [\href{https://arxiv.org/abs/1002.4405}{{\tt
  1002.4405}}].

\bibitem{Hisano05}
J.~{Hisano}, S.~{Matsumoto}, M.~M. {Nojiri} and O.~{Saito},
  \emph{{Nonperturbative effect on dark matter annihilation and gamma ray
  signature from the galactic center}},
  \href{http://dx.doi.org/10.1103/PhysRevD.71.063528}{\emph{\prd} {\bf 71}
  (Mar., 2005) 063528}, [\href{https://arxiv.org/abs/hep-ph/0412403}{{\tt
  hep-ph/0412403}}].

\bibitem{AHDM}
N.~{Arkani-Hamed}, D.~P. {Finkbeiner}, T.~R. {Slatyer} and N.~{Weiner},
  \emph{{A theory of dark matter}},
  \href{http://dx.doi.org/10.1103/PhysRevD.79.015014}{\emph{\prd} {\bf 79}
  (Jan., 2009) 015014}, [\href{https://arxiv.org/abs/0810.0713}{{\tt
  0810.0713}}].

\bibitem{Shepherd09}
W.~{Shepherd}, T.~M.~P. {Tait} and G.~{Zaharijas}, \emph{{Bound states of
  weakly interacting dark matter}},
  \href{http://dx.doi.org/10.1103/PhysRevD.79.055022}{\emph{\prd} {\bf 79}
  (Mar., 2009) 055022}, [\href{https://arxiv.org/abs/0901.2125}{{\tt
  0901.2125}}].

\bibitem{Lattanzi2009}
M.~{Lattanzi} and J.~{Silk}, \emph{{Can the WIMP annihilation boost factor be
  boosted by the Sommerfeld enhancement?}},
  \href{http://dx.doi.org/10.1103/PhysRevD.79.083523}{\emph{Phys. Rev. D} {\bf
  79} (Apr., 2009) 083523}, [\href{https://arxiv.org/abs/0812.0360}{{\tt
  0812.0360}}].

\bibitem{Hryczuk11}
A.~{Hryczuk}, R.~{Iengo} and P.~{Ullio}, \emph{{Relic densities including
  Sommerfeld enhancements in the MSSM}},
  \href{http://dx.doi.org/10.1007/JHEP03(2011)069}{\emph{\jhep} {\bf 3} (Mar.,
  2011) 69}, [\href{https://arxiv.org/abs/1010.2172}{{\tt 1010.2172}}].

\bibitem{Ibe2009}
M.~{Ibe}, H.~{Murayama} and T.~T. {Yanagida}, \emph{{Breit-Wigner enhancement
  of dark matter annihilation}},
  \href{http://dx.doi.org/10.1103/PhysRevD.79.095009}{\emph{Phys. Rev. D} {\bf
  79} (May, 2009) 095009}, [\href{https://arxiv.org/abs/0812.0072}{{\tt
  0812.0072}}].

\bibitem{Strigari2007}
L.~E. {Strigari}, S.~M. {Koushiappas}, J.~S. {Bullock} and M.~{Kaplinghat},
  \emph{{Precise constraints on the dark matter content of MilkyWay dwarf
  galaxies for gamma-ray experiments}},
  \href{http://dx.doi.org/10.1103/PhysRevD.75.083526}{\emph{\prd} {\bf 75}
  (Apr., 2007) 083526}, [\href{https://arxiv.org/abs/astro-ph/0611925}{{\tt
  astro-ph/0611925}}].

\bibitem{Kuhlen2008}
M.~{Kuhlen}, J.~{Diemand} and P.~{Madau}, \emph{{The Dark Matter Annihilation
  Signal from Galactic Substructure: Predictions for GLAST}},
  \href{http://dx.doi.org/10.1086/590337}{\emph{ApJ} {\bf 686} (Oct., 2008)
  262--278}, [\href{https://arxiv.org/abs/0805.4416}{{\tt 0805.4416}}].

\bibitem{Elahi2009}
P.~J. {Elahi}, L.~M. {Widrow} and R.~J. {Thacker}, \emph{{Can substructure in
  the Galactic halo explain the ATIC and PAMELA results?}},
  \href{http://dx.doi.org/10.1103/PhysRevD.80.123513}{\emph{Phys. Rev. D} {\bf
  80} (Dec., 2009) 123513}, [\href{https://arxiv.org/abs/0906.4352}{{\tt
  0906.4352}}].

\bibitem{Scott2009}
P.~{Scott} and S.~{Sivertsson}, \emph{{Gamma Rays from Ultracompact Primordial
  Dark Matter Minihalos}},
  \href{http://dx.doi.org/10.1103/PhysRevLett.103.211301}{\emph{Phys. Rev.
  Lett.} {\bf 103} (Nov., 2009) 211301},
  [\href{https://arxiv.org/abs/0908.4082}{{\tt 0908.4082}}].

\bibitem{Kamionkowski2010}
M.~{Kamionkowski}, S.~M. {Koushiappas} and M.~{Kuhlen}, \emph{{Galactic
  substructure and dark-matter annihilation in the Milky Way halo}},
  \href{http://dx.doi.org/10.1103/PhysRevD.81.043532}{\emph{Phys. Rev. D} {\bf
  81} (Feb., 2010) 043532}, [\href{https://arxiv.org/abs/1001.3144}{{\tt
  1001.3144}}].

\bibitem{Anderhalden2013}
D.~{Anderhalden} and J.~{Diemand}, \emph{{Density profiles of CDM microhalos
  and their implications for annihilation boost factors}},
  \href{http://dx.doi.org/10.1088/1475-7516/2013/04/009}{\emph{JCAP} {\bf 4}
  (Apr., 2013) 009}, [\href{https://arxiv.org/abs/1302.0003}{{\tt 1302.0003}}].

\bibitem{Bartels2015}
R.~{Bartels} and S.~{Ando}, \emph{{Boosting the annihilation boost: Tidal
  effects on dark matter subhalos and consistent luminosity modeling}},
  \href{http://dx.doi.org/10.1103/PhysRevD.92.123508}{\emph{Phys. Rev. D} {\bf
  92} (Dec., 2015) 123508}, [\href{https://arxiv.org/abs/1507.08656}{{\tt
  1507.08656}}].

\bibitem{Stref2016}
M.~{Stref} and J.~{Lavalle}, \emph{{Modeling dark matter subhalos in a
  constrained galaxy: Global mass and boosted annihilation profiles}},
  {\emph{ArXiv e-prints} (Oct., 2016) },
  [\href{https://arxiv.org/abs/1610.02233}{{\tt 1610.02233}}].

\bibitem{Mack2014}
K.~J. {Mack}, \emph{{Known unknowns of dark matter annihilation over cosmic
  time}}, \href{http://dx.doi.org/10.1093/mnras/stu129}{\emph{\mnras} {\bf 439}
  (Apr., 2014) 2728--2735}, [\href{https://arxiv.org/abs/1309.7783}{{\tt
  1309.7783}}].

\bibitem{Sanchez-Conde2014}
M.~A. {S{\'a}nchez-Conde} and F.~{Prada}, \emph{{The flattening of the
  concentration-mass relation towards low halo masses and its implications for
  the annihilation signal boost}},
  \href{http://dx.doi.org/10.1093/mnras/stu1014}{\emph{MNRAS} {\bf 442} (Aug.,
  2014) 2271--2277}, [\href{https://arxiv.org/abs/1312.1729}{{\tt 1312.1729}}].

\bibitem{Berezinsky2003}
V.~{Berezinsky}, V.~{Dokuchaev} and Y.~{Eroshenko}, \emph{{Small-scale clumps
  in the galactic halo and dark matter annihilation}},
  \href{http://dx.doi.org/10.1103/PhysRevD.68.103003}{\emph{Phys. Rev. D} {\bf
  68} (Nov., 2003) 103003}, [\href{https://arxiv.org/abs/astro-ph/0301551}{{\tt
  astro-ph/0301551}}].

\bibitem{Ricotti2009}
M.~{Ricotti} and A.~{Gould}, \emph{{A New Probe of Dark Matter and High-Energy
  Universe Using Microlensing}},
  \href{http://dx.doi.org/10.1088/0004-637X/707/2/979}{\emph{ApJ} {\bf 707}
  (Dec., 2009) 979--987}, [\href{https://arxiv.org/abs/0908.0735}{{\tt
  0908.0735}}].

\bibitem{Bertschinger1985}
E.~{Bertschinger}, \emph{{Self-similar secondary infall and accretion in an
  Einstein-de Sitter universe}},
  \href{http://dx.doi.org/10.1086/191028}{\emph{ApJS} {\bf 58} (May, 1985)
  39--65}.

\bibitem{1985davis}
M.~{Davis}, G.~{Efstathiou}, C.~S. {Frenk} and S.~D.~M. {White}, \emph{{The
  evolution of large-scale structure in a universe dominated by cold dark
  matter}}, \href{http://dx.doi.org/10.1086/163168}{\emph{\apj} {\bf 292} (May,
  1985) 371--394}.

\bibitem{2005natureSpringel}
V.~{Springel}, S.~D.~M. {White}, A.~{Jenkins}, C.~S. {Frenk}, N.~{Yoshida},
  L.~{Gao} et~al., \emph{{Simulations of the formation, evolution and
  clustering of galaxies and quasars}},
  \href{http://dx.doi.org/10.1038/nature03597}{\emph{\nat} {\bf 435} (June,
  2005) 629--636}, [\href{https://arxiv.org/abs/astro-ph/0504097}{{\tt
  astro-ph/0504097}}].

\bibitem{Bringmann2012}
T.~{Bringmann}, P.~{Scott} and Y.~{Akrami}, \emph{{Improved constraints on the
  primordial power spectrum at small scales from ultracompact minihalos}},
  \href{http://dx.doi.org/10.1103/PhysRevD.85.125027}{\emph{Phys. Rev. D} {\bf
  85} (June, 2012) 125027}, [\href{https://arxiv.org/abs/1110.2484}{{\tt
  1110.2484}}].

\bibitem{Berezinsky2012}
V.~S. {Berezinsky}, V.~I. {Dokuchaev} and Y.~N. {Eroshenko}, \emph{{Formation
  of superdense dark matter lumps at the radiation-dominated cosmological
  stage}}, \href{http://dx.doi.org/10.1134/S0202289312010045}{\emph{Gravitation
  and Cosmology} {\bf 18} (Jan., 2012) 57--60}.

\bibitem{Berezinsky2013}
V.~S. {Berezinsky}, V.~I. {Dokuchaev} and Y.~N. {Eroshenko}, \emph{{Formation
  and internal structure of superdense dark matter clumps and ultracompact
  minihaloes}},
  \href{http://dx.doi.org/10.1088/1475-7516/2013/11/059}{\emph{JCAP} {\bf 11}
  (Nov., 2013) 59}, [\href{https://arxiv.org/abs/1308.6742}{{\tt 1308.6742}}].

\bibitem{Navarro1996}
J.~F. {Navarro}, C.~S. {Frenk} and S.~D.~M. {White}, \emph{{The Structure of
  Cold Dark Matter Halos}}, \href{http://dx.doi.org/10.1086/177173}{\emph{ApJ}
  {\bf 462} (May, 1996) 563},
  [\href{https://arxiv.org/abs/astro-ph/9508025}{{\tt astro-ph/9508025}}].

\bibitem{Zhang2011}
D.~{Zhang}, \emph{{Impact of primordial ultracompact minihaloes on the
  intergalactic medium and first structure formation}},
  \href{http://dx.doi.org/10.1111/j.1365-2966.2011.19602.x}{\emph{\mnras} {\bf
  418} (Dec., 2011) 1850--1872}, [\href{https://arxiv.org/abs/1011.1935}{{\tt
  1011.1935}}].

\bibitem{Yang2016}
Y.~{Yang}, \emph{{Contributions of dark matter annihilation within ultracompact
  minihalos to the 21 cm background signal}},
  \href{http://dx.doi.org/10.1140/epjp/i2016-16432-8}{\emph{European Physical
  Journal Plus} {\bf 131} (Dec., 2016) 432},
  [\href{https://arxiv.org/abs/1612.06559}{{\tt 1612.06559}}].

\bibitem{Bringmann2009}
T.~{Bringmann}, \emph{{Particle models and the small-scale structure of dark
  matter}}, \href{http://dx.doi.org/10.1088/1367-2630/11/10/105027}{\emph{New
  J. Phys.} {\bf 11} (Oct., 2009) 105027},
  [\href{https://arxiv.org/abs/0903.0189}{{\tt 0903.0189}}].

\bibitem{Berezinsky2006}
V.~{Berezinsky}, V.~{Dokuchaev} and Y.~{Eroshenko}, \emph{{Destruction of
  small-scale dark matter clumps in the hierarchical structures and galaxies}},
  \href{http://dx.doi.org/10.1103/PhysRevD.73.063504}{\emph{Phys. Rev. D} {\bf
  73} (Mar., 2006) 063504}, [\href{https://arxiv.org/abs/astro-ph/0511494}{{\tt
  astro-ph/0511494}}].

\bibitem{Berezinsky2008}
V.~{Berezinsky}, V.~{Dokuchaev} and Y.~{Eroshenko}, \emph{{Remnants of dark
  matter clumps}},
  \href{http://dx.doi.org/10.1103/PhysRevD.77.083519}{\emph{Phys. Rev. D} {\bf
  77} (Apr., 2008) 083519}, [\href{https://arxiv.org/abs/0712.3499}{{\tt
  0712.3499}}].

\bibitem{Sutherland1993}
R.~S. {Sutherland} and M.~A. {Dopita}, \emph{{Cooling functions for low-density
  astrophysical plasmas}}, \href{http://dx.doi.org/10.1086/191823}{\emph{\apjs}
  {\bf 88} (Sept., 1993) 253--327}.

\bibitem{Springel2005a}
V.~{Springel}, \emph{{The cosmological simulation code GADGET-2}},
  \href{http://dx.doi.org/10.1111/j.1365-2966.2005.09655.x}{\emph{MNRAS} {\bf
  364} (Dec., 2005) 1105--1134},
  [\href{https://arxiv.org/abs/astro-ph/0505010}{{\tt astro-ph/0505010}}].

\bibitem{Hooper2007}
D.~{Hooper}, D.~P. {Finkbeiner} and G.~{Dobler}, \emph{{Possible evidence for
  dark matter annihilations from the excess microwave emission around the
  center of the Galaxy seen by the Wilkinson Microwave Anisotropy Probe}},
  \href{http://dx.doi.org/10.1103/PhysRevD.76.083012}{\emph{\prd} {\bf 76}
  (Oct., 2007) 083012}, [\href{https://arxiv.org/abs/0705.3655}{{\tt
  0705.3655}}].

\bibitem{Hooper2011}
D.~{Hooper} and T.~{Linden}, \emph{{Gamma rays from the Galactic center and the
  WMAP haze}}, \href{http://dx.doi.org/10.1103/PhysRevD.83.083517}{\emph{\prd}
  {\bf 83} (Apr., 2011) 083517}, [\href{https://arxiv.org/abs/1011.4520}{{\tt
  1011.4520}}].

\bibitem{Cholis2015}
I.~{Cholis}, D.~{Hooper} and T.~{Linden}, \emph{{A critical reevaluation of
  radio constraints on annihilating dark matter}},
  \href{http://dx.doi.org/10.1103/PhysRevD.91.083507}{\emph{\prd} {\bf 91}
  (Apr., 2015) 083507}, [\href{https://arxiv.org/abs/1408.6224}{{\tt
  1408.6224}}].

\bibitem{Delahaye2010}
T.~{Delahaye}, J.~{Lavalle}, R.~{Lineros}, F.~{Donato} and N.~{Fornengo},
  \emph{{Galactic electrons and positrons at the Earth: new estimate of the
  primary and secondary fluxes}},
  \href{http://dx.doi.org/10.1051/0004-6361/201014225}{\emph{\aap} {\bf 524}
  (Dec., 2010) A51}, [\href{https://arxiv.org/abs/1002.1910}{{\tt 1002.1910}}].

\bibitem{SPF}
T.~R. {Slatyer}, N.~{Padmanabhan} and D.~P. {Finkbeiner}, \emph{{CMB
  constraints on WIMP annihilation: Energy absorption during the recombination
  epoch}}, \href{http://dx.doi.org/10.1103/PhysRevD.80.043526}{\emph{Phys. Rev.
  D} {\bf 80} (Aug., 2009) 043526},
  [\href{https://arxiv.org/abs/0906.1197}{{\tt 0906.1197}}].

\bibitem{Slatyer2013}
T.~R. {Slatyer}, \emph{{Energy injection and absorption in the cosmic dark
  ages}}, \href{http://dx.doi.org/10.1103/PhysRevD.87.123513}{\emph{Phys. Rev.
  D} {\bf 87} (June, 2013) 123513},
  [\href{https://arxiv.org/abs/1211.0283}{{\tt 1211.0283}}].

\bibitem{galactics1}
K.~{Kuijken} and J.~{Dubinski}, \emph{{Nearly Self-Consistent Disc / Bulge /
  Halo Models for Galaxies}},
  \href{http://dx.doi.org/10.1093/mnras/277.4.1341}{\emph{MNRAS} {\bf 277}
  (Dec., 1995) 1341}.

\bibitem{galactics2}
L.~M. {Widrow} and J.~{Dubinski}, \emph{{Equilibrium Disk-Bulge-Halo Models for
  the Milky Way and Andromeda Galaxies}},
  \href{http://dx.doi.org/10.1086/432710}{\emph{ApJ} {\bf 631} (Oct., 2005)
  838--855}, [\href{https://arxiv.org/abs/astro-ph/0506177}{{\tt
  astro-ph/0506177}}].

\bibitem{galactics3}
L.~M. {Widrow}, B.~{Pym} and J.~{Dubinski}, \emph{{Dynamical Blueprints for
  Galaxies}}, \href{http://dx.doi.org/10.1086/587636}{\emph{ApJ} {\bf 679}
  (June, 2008) 1239--1259}, [\href{https://arxiv.org/abs/0801.3414}{{\tt
  0801.3414}}].

\bibitem{pynbody}
A.~{Pontzen}, R.~{Ro{\v s}kar}, G.~{Stinson} and R.~{Woods}, ``{pynbody:
  N-Body/SPH analysis for python}.'' Astrophysics Source Code Library, May,
  2013.

\bibitem{JG10}
A.~S. {Josan} and A.~M. {Green}, \emph{{Gamma rays from ultracompact minihalos:
  Potential constraints on the primordial curvature perturbation}},
  \href{http://dx.doi.org/10.1103/PhysRevD.82.083527}{\emph{\prd} {\bf 82}
  (Oct., 2010) 083527}, [\href{https://arxiv.org/abs/1006.4970}{{\tt
  1006.4970}}].

\bibitem{Shandera12}
S.~{Shandera}, A.~L. {Erickcek}, P.~{Scott} and J.~Y. {Galarza}, \emph{{Number
  counts and non-Gaussianity}},
  \href{http://dx.doi.org/10.1103/PhysRevD.88.103506}{\emph{\prd} {\bf 88}
  (Nov., 2013) 103506}, [\href{https://arxiv.org/abs/1211.7361}{{\tt
  1211.7361}}].

\bibitem{Clark2016}
H.~A. {Clark}, G.~F. {Lewis} and P.~{Scott}, \emph{{Investigating dark matter
  substructure with pulsar timing - II. Improved limits on small-scale
  cosmology}}, \href{http://dx.doi.org/10.1093/mnras/stv2529}{\emph{\mnras}
  {\bf 456} (Feb., 2016) 1402--1409},
  [\href{https://arxiv.org/abs/1509.02941}{{\tt 1509.02941}}].

\bibitem{Aslanyan16}
G.~{Aslanyan}, L.~C. {Price}, J.~{Adams}, T.~{Bringmann}, H.~A. {Clark},
  R.~{Easther} et~al., \emph{{Ultracompact Minihalos as Probes of Inflationary
  Cosmology}},
  \href{http://dx.doi.org/10.1103/PhysRevLett.117.141102}{\emph{\prl} {\bf 117}
  (Sept., 2016) 141102}, [\href{https://arxiv.org/abs/1512.04597}{{\tt
  1512.04597}}].

\end{thebibliography}\endgroup
\end{document}